# FALACO SOLITONS
## Cosmic strings in a swimming pool

**R. M. Kiehn**
69 Chemin St. Donat, 84380 Mazan, France
Http://www.cartan.pair.com
Rkiehn2352@aol.com

**Abstract**: Topological defects experimentally induced by rotational dynamics in a continuous media replicate the coherent structure features of cosmic strings as well as hadrons.

**Introduction:**

During the summer of 1986, while visiting an old friend in Rio de Janeiro, Brazil, the present author became aware of a significant topological experiment that can be replicated by almost everyone with access to a swimming pool. See Photo 1. In short, it is possible to produce, hydrodynamically, in a surface of discontinuity, a long lived coherent structure that consists of a set of macroscopic topological defects. This long-lived coherent object, dubbed the Falaco Soliton, has several features equivalent to those reported for models of the submicroscopic hadron. String theorists take note, for the structure consists of a pair of topological 2-dimensional rotational defects in a surface of discontinuity, globally connected and stabilized by a 1 dimensional topological defect or string.

This experimental observation greatly stimulated this author to further research in applied topology involving topological defects, and the topological evolution of such defects which can be associated with phase changes and thermodynamically irreversible and turbulent phenomena [1],[6]. When colleagues would ask "What is a topological defect?" it was possible to point to something that they could replicate and understand visually at a macroscopic level. The original observation was first described at a Dynamics Days conference in Austin, TX, [2] and has been reported, as parts of other research, in various hydrodynamic publications [3], but it is apparent that these concepts have not penetrated into other areas of research. As the phenomena is a topological issue, and can happen at all scales, the Falaco Soliton should be a natural artifact of both the sub-atomic and the cosmological worlds. The reason d'etre for this short article is to bring the idea to the attention of other researchers who might find the concept interesting and stimulating to their own research.

**The Experiment**

The phenomena is easily reproduced by placing a large half-submerged circular disc or plate (a Frisbee will do) in a swimming pool, then stroking the plate slowly in the direction of its oblate axis. At the end of the stroke, slowly and smoothly extract the plate from the water, imparting kinetic energy and distributed angular momentum to the fluid. Initially, the edges of the plate will create a pair of Rankine Vortices in the surface of the water (a density discontinuity). These Rankine vortices cause the initially flat surface of discontinuity to form a pair of parabolic concave indentations of positive Gauss curvature, indicative of the "rigid body" rotation of a pair of contra-rotating vortex cores of uniform vorticity.   In a few seconds the concave Rankine depressions, with visible spiral arm caustics, will decay into a pair of convex dimples of negative Gauss curvature.  The surface effects can be observed in bright sunlight via their Snell projections as large black spots on the bottom of the pool. In a few tries you will become an expert experimentalist, for the drifting spots are easily created and surprisingly will persist for many minutes in a still pool.

The dimpled depressions are typically of the order of a few millimeters, but the zone of circulation typically extends over a disc of some 10 to 25 centimeters or more, depending on the plate diameter. This configuration, or coherent structure, has been defined as the Falaco Soliton. See Figure 1. For purposes of illustration , the vertical depression has been greatly exaggerated.

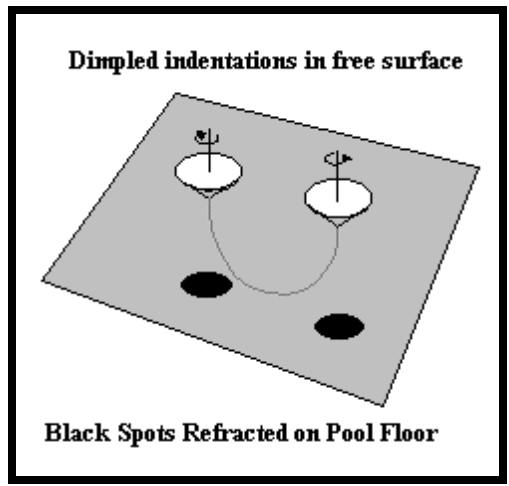

**Figure 1. The Falaco Soliton**

The effect is easily observed, for in strong sunlight the convex hyperbolic indentation will cause an intensely black circular disk (or absence of light) to be imaged on the bottom of the pool. In addition a bright ring of focused light will surround the black disk, emphasizing the contrast. During the initial few seconds of decay to the meta-stable soliton state, the large black disk is decorated with spiral arm caustics, remindful of spiral arm galaxies. The spiral arm caustics contract around the large black disk during the stabilization process, and ultimately disappear when the soliton state is achieved. It should be noted that if chalk dust is sprinkled on the surface of the pool during the formative stages of the Falaco soliton, then the topological signature of the familiar Mushroom Spiral pattern is exposed.

The black disk optics are completely described by Snell refraction from a surface of revolution that has negative Gauss curvature. See Figure 2. This effect has been reported elsewhere [2], but the figures are replicated herein for clarity.

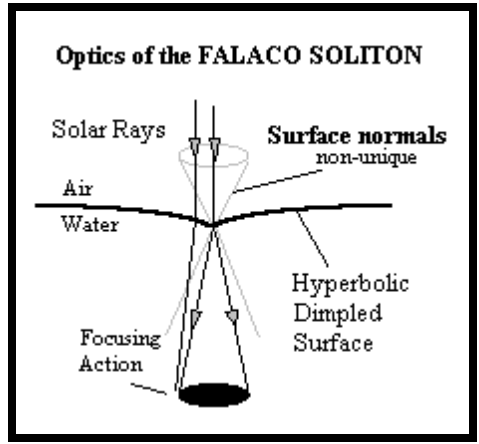

**Figure 2. The Snell Black Disk**

Early on, it was recognized that the Snell refraction on the bottom of the pool produces a circular disk, more or less independent from the angle of solar incidence. This observation, as well as the negative Gauss curvature of the surface, lends further credence to the idea that the topological surface distortion is a *minimal* surface. The argument is that only spheres and minimal surfaces have a conformal Gauss map (Snell projection) which projects circles into circles. The long life of the soliton state in the presence of a viscous media indicates that the flow vector field describing the dynamics is harmonic, and generates a conformal minimal surface in space time. Guided by the fact that all holomorphic functions in four dimensions generate minimal surfaces, an approximate representation is given by the harmonic 4-vector field

$$W(x,y,z,t) = [\exp(-kr)\cos(\omega t), \exp(-kr)\sin(\omega t), -kr, \omega t].$$

The associated minimal surface has circular symmetry, and a hyperbolic profile similar to the observed dimple shape. Though created from a time dependent field, the minimal surface profile generated by the above formula is constant in time. However, a full solution describing the experimental effect has not been found. It would seem that string theorists should be able to find a reasonable solution to this dynamical system if their theories have any semblence of being correct.

Recently it has been recognized that a complex 2 surface induced by the Hopf map can have positive Gauss curvature and yet be a minimal surface. The surface curvatures are pure imaginary and of opposite sign. It is conjectured that the initial state of Rankine vortices might be such a minimal surface of positive Gauss curvature which dynamically undergoes a topological phase change to form the negative Gauss curvature minimal surface of the final Falaco state.

Dye injection near an axis of rotation during the formative stages indicates that there is a unseen thread, or 1-dimensional string singularity, in the form of a circular arc that connects the two 2-dimensional surface singularities or dimples. *Transverse* waves of dye streaks can be observed to propagate, back and forth, from one dimple vertex to the other dimple vertex, guided by the "string" singularity. The effect is remindful of the whistler propagation of electrons along the guiding center of the earth's magnetic field lines.

If a thin broom handle or a rod is placed vertically in the pool, and the Falaco soliton pair is directed in its translation motion to intercept the rod symmetrically, as the soliton pair comes within of the "scattering center - soliton" range (approximately the separation distance of the two rotation centers) the large black spots at first shimmer and disappear. Then a short time later, after the soliton has passed beyond the interaction range of the scattering center, the large black spots coherently reappear, mimicking the numerical simulations of soliton scattering. For hydrodynamics, this observation firmly cements the idea that these objects are truly coherent structures.

If the string is sharply "severed", the confined, two dimensional endcap singularities do not diffuse away, but instead disappear almost explosively. It is this observation that leads to the statement that the Falaco soliton is the macroscopic topological equivalent of the illusive hadron in elementary particle theory. The two 2-dimensional surface defects (the quarks) are bound together by a string of confinement, and cannot be isolated. The dynamics of such a coherent structure is extraordinary, for it is a system that is globally stabilized by the presence of the connecting 1-dimensional string.

**Summary**

As the Falaco phenomena appears to be the result of a topological defect, it follows that as a topological property of hydrodynamic evolution, it could appear in any density discontinuity, at any scale. This spin pairing mechanism, as a topological phenomenon, is independent from size and shape, and could occur at both the microscopic and the cosmic scales. In fact, as mentioned above, during the formative stages of the Falaco Soliton pair, the decaying Rankine vortices exhibit spiral arms easily visible as caustics emanating from the boundary of each vortex core. The observation is so striking that it leads to the conjecture: "Can the nucleus of M31 be connected to the nucleus of our Milky way galaxy by a tubular cosmic thread?

At smaller scales, the concept also permits the development of another mechanism for producing spin-pairing of electrons in the discontinuity of the Fermi surface, or in two dimensional charge distributions. Could this spin pairing mechanism, depending on transverse wave, not longitudinal wave, coupling be another mechanism for explaining superconductivity? As the defect is inherently 3-dimensional, it must be associated with a 3-form of Topological Torsion, $A^\wedge dA$, introduced by the author ian 1976 [4], but now more commonly called a Chern Simons term.. These ideas were exploited in an attempt to explain high TC superconductivity. [5]

To this author the importance of the Falaco Solitons is that they offer the first clean experimental evidence of topological defects taking place in a dynamical system.

Moreover, the experiments are fascinating, easily replicated by anyone with access to a swimming pool, and stimulate thinking in almost everyone that observes them, no matter what his field of expertise. They certainly are among the most easily produced solitons.

**Acknowledgements**

**References**
[ 1] R. M. Kiehn, (1997) "Coherent Structures in Fluids are Deformable Topological Torsion Defects" presented at the IUTAM-SIMFLO Conference at DTU, Denmark, May 25-29, 1997.
[2] R. M. Kiehn, Talk at 1987 Dynamics Days, Austin, Texas
[3] J. Sterling et.al., Phys Fluids **30** 11, 1987.
....Kiehn, R. M., 1991, "Compact Dissipative Flow Structures with Topological Coherence Embedded in Eulerian Environments", in: *Non-linear Dynamics of Structures*, edited by R.Z. Sagdeev, U. Frisch, F. Hussain, S. S. Moiseev and N. S. Erokhin, (World Scientific Press, Singapore ) p.139-164.
.... (1992) "Topological Defects, Coherent Structures and Turbulence in Terms of Cartan's Theory of Differential Topology" in Developments in Theoretical and Applied Mathematics, Proceedings of the SECTAM XVI conference, B. N. Antar, R. Engels, A.A. Prinaris and T. H. Moulden, Editors, The University of Tennessee Space Institute, Tullahoma, TN 37388 USA. p.III.IV.02
.... (1995) "Hydrodynamic Wakes and Minimal Surfaces with Fractal Boundaries" in Mixing in Geophysical Flows, J. M. Redondo and O. Metais, Editors, CIMNE Barcelona, Spain.
[4] R. M. Kiehn, NASA AMES NCA-2-OR-295-502 (1976)
[5] Kiehn, R. M., "Are there three kinds of superconductivity", Int. J. of Mod. Phys. **1**0,1779 (1991)
[6] http://www.cartan.pair.com

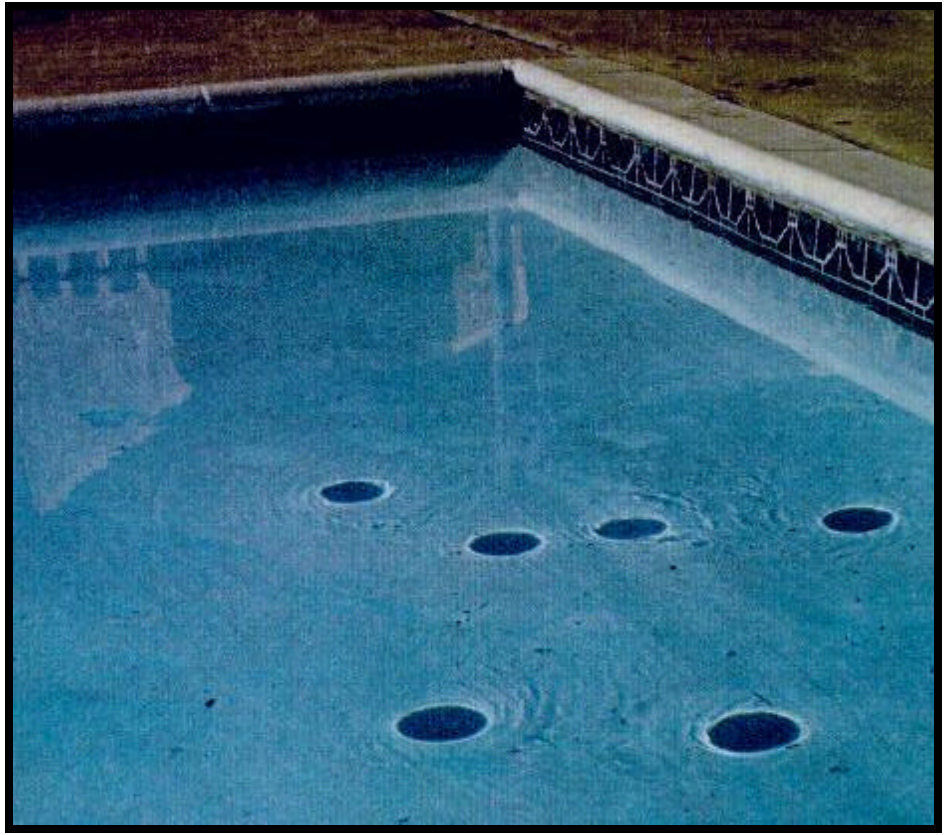

**Photo 1.** Three pairs of FALACO SOLITONS, a few minutes after creation. The kinetic energy and the angular momentum of a pair of Rankine vortices created in the free surface of water quickly decay into dimpled, locally unstable, singular surfaces that have an extraordinary lifetime of many minutes in a still pool. These singular surfaces are connected by means of a stabilizing invisible singular thread, or string, which if abruptly severed will cause the endcaps to disappear in a rapid non-diffusive manner. The black discs are formed on the bottom of the pool by Snell refraction of a rotationally induced dimpled surface. Careful examination of contrast in the photo will indicate the region of the dimpled surface as artifacts to the left of each black spot at a distance about equal to the separation distance of the top right pair and elevated above the horizon by about 25 degrees. The photo was taken in late afternoon. The fact that the projections are circular and not ellipses indicates that the dimpled surface is a minimal surface.
(Photo by David Radabaugh, Schlumberger, Houston)